\documentclass[12pt]{article}
\usepackage{graphics}
\begin{document}
\markright{Nonquantum Gravity}
\title{Nonquantum Gravity}

\author{Stephen Boughn
{\small\thanks{sboughn@haverford.edu}}
\\[2mm]
Departments of Physics and Astronomy \\
Haverford College, Haverford, PA 19041 }

\date{{\small   \LaTeX-ed \today}}

\maketitle

\begin{abstract}
One of the great challenges for 21st century physics is to quantize 
gravity and generate a theory that will unify gravity with the other three 
fundamental forces of nature.  This paper takes the (heretical) point of 
view that gravity may be an inherently classical, i.e., nonquantum, phenomenon
and investigates the experimental consequences of such a conjecture.  At 
present there is no experimental evidence of the quantum nature of gravity 
and the likelihood of definitive tests in the future is not at all certain. 
If gravity is, indeed, a nonquantum phenomenon, then it is suggested that
evidence will most likely appear at mesoscopic scales.

\vspace*{5mm} \noindent PACS: .03.65.Ta, .04.80.Cc, 95.30.Sf
\\ Keywords: Gravity, Quantum Theory.

\end{abstract}

\section{Introduction}
\label{intro}

Ever since the earliest attempts by Einstein and others \cite{wey18,kal21,edd23} to unify the 
two fundamental fields of classical physics, gravity and electromagnetism, one
of the major goals of physics has been to unify all the fundamental 
forces of nature.  Such a unification was not without precedent; after
all, Maxwell had previously unified electricity and magnetism.  Since then,
the electromagnetic and weak forces have been successfully unified (electroweak theory)
and these have been, more or less, unified with the strong force in the Standard Model 
of particle physics.  So far, unifying gravity with the strong, weak, and 
electromagnetic forces has proved to be much more difficult and only minor progress
has been made; although, the proponents of superstring theory have great hopes that 
they are on the right track.  It seems reasonable to expect that in order to unify 
gravity with the other quantum theoretical forces, gravity must also be quantum
mechanical in origin.  Efforts to quantize gravity began shortly after the advent of 
quantum mechanics and continue to this day.  While there are several intriquing candidates, 
including superstring theory \cite{gre87} and loop quantum gravity \cite{ash92} 
there is, as of yet, no successful 
theory of quantum gravity.  Quantizing gravity is, to be sure, fraught with technical
difficulties but the total lack of any observations of the quantum nature of gravity must 
certainly be listed as one of the important reasons for lack of progress.

Because there is currently absolutely no experimental evidence of the quantum nature of
gravity, it seems reasonable to at least contemplate the hypothesis that gravity is an
entirely classical, nonquantum phenomenon.  While such a conjecture is, perhaps, viewed 
as heretical by most physicists, it is not a new conjecture \cite{mol62,ros63,car08}.  The 
study of the possibility of coupling quantum and classical systems is an active area of 
research motivated primarily by ``the measurement problem'' in quantum mechanics, i.e.,
how it is that a quantum mechanical system can interact with a measuring apparatus to
yield a classical observation \cite{zur03,hal05}.  Most of this research is theoretical 
in nature and much of it involves the theoretical consistency of a classical-quantum 
coupling.  The present paper, by contrast, is motivated almost entirely by what has and 
what has not been observed or by what can and what cannot be experimentally verified.  
In short, it is the analysis of an experimental physicist (which I am).   Little 
attention is paid to theoretical consistency, for example, even whether or not energy
and momentum are conserved {\it if} there is no observable consequence of such an
inconsistency.  For this reason, 
the conjectures of this paper by no means constitute a theory or even a simple model of 
gravity but are offered simply to entice others to look seriously at the 
possibility of nonquantum gravity. 

 In one model of nonquantum gravity, semi-classical gravity,  Einstein's equations
\begin{equation}
G_{\mu \nu} = \frac{8\pi G}{c^4} T_{\mu \nu}
\label{einstein}
\end{equation}
are replaced with
\begin{equation}
G_{\mu \nu} = \frac{8\pi G}{c^4} \langle T_{\mu \nu} \rangle
\label{einstein2}
\end{equation}
where $\langle T_{\mu \nu} \rangle$ is the expection value of the stress-energy
operator \cite{car08}.  There have been both experimental and theoretical objections 
to this generalization and these will be discussed in \S \ref{consistency}.  The primary 
conjecture of this paper is that the classical Einstein's equations of Eq. \ref{einstein}
describe gravity where the stress-energy tensor, $T_{\mu \nu}$, has its usual 
classical meaning.  Of course, this classical description is possible only when the
classical stress-energy of the system is well-defined.  A prerequisite for this is,
in the language of decoherence theory, that the system is in a decoherent, mixed 
quantum state for only then do the probability predictions of quantum theory 
agree with those of classical physics.  Even then, in the fully classical case, there is a single 
stress-energy tensor while in the general decoherent quantum state the expectation value of the
stress-energy operator is a linear combination of possible stress-energy tensors.  
This distinction is crucial because the gravitational fields implied by Eqs. \ref{einstein}
and \ref{einstein2} are not the same.

In cases where the source is in a coherent, non-localized quantum state, there
is no well-defined, classical stress-energy tensor and Eq. \ref{einstein} is of no use
in specifying the gravitational field.  (Just what non-localized and coherent mean in this 
context will be dealt with in \S \ref{decoherence}.)  It is tempting to conjecture that such systems 
are not sources of gravity; however, it is not necessary to be this unequivocal.
It is the conjecture of  this paper that the gravitational field of a non-localized, 
coherent quantum system is simply not well-defined. Therefore, Eq. \ref{einstein}
need not be invoked and the above quandry does not arise.  It will turn 
out that for microsopic systems, in which quantum coherence is most commonly observed, the 
effects of gravity are, in principle, unobservable.  For larger macroscopic systems,
decoherence is assured and the classical stress-energy is well-defined.
This leaves open the question of gravitational interactions of mesoscopic, coherent
systems.  It is these and other issues that are discussed below. 

The following section deals with the detectability or rather undetectability of gravitons, 
which was one motivation for the present paper. \S \ref{interference} investigates the conditions 
under which coherent quantum systems can have measureable gravitational effects and an
implied mass lower limit required to observe these effects.  \S \ref{GWs} concerns the emission
of gravitational radiation and whether or not this process affects quantum coherence.
\S \ref{decoherence} deals with the effects of gravity in mesoscopic systems and the relation to 
quantum decoherence.  Inherent inconsistencies in a nonquantum theory of gravity and 
previous claims that gravity must be quantized are discussed in \S \ref{consistency}.  
\S \ref{discussion} includes a brief discussion of the importance of classicality in quantum
measurements as well as further speculation on the consequences of nonquantum gravity.

I note in passing that another approach to unifying gravity with the other fundamental forces 
of nature is to consider gravity as an emergent mean field approximation of more fundamental 
underlying microscopic phenomena.  Some such theories have already been ruled out 
\cite{wei80} while others are still under consideration \cite{bar01}.  However, I will not 
discuss any relation between emergent theories and the nonquantum conjecture of this paper 
other than to note that the no-go theorem of Weinberg and Witten \cite{wei80} is not
applicable because an implicit conjecture of this paper is that gravitons do not exist.

\section{The Detectability of Gravitons}
\setcounter{equation}{0} \label{graviton}

If gravity is, indeed, a quantum phenomenon, it seems quite likely that a prominent 
feature of the theory will be the graviton, the fundamental quantum of gravitational
radiation with energy $h \nu$ where $\nu$ is the frequency of the radiation.
Several years ago, Freeman Dyson posed the question of whether or not gravitons
can in principle be detected, that is, whether the quantization of the gravitational
field can ever be detected \cite{dys04}.  If not, Dyson continues, then one wonders
whether or not gravitons actually exist.   Motivated by Dyson's question, Tony Rothman 
and I investigated the detectability of gravitons \cite{rot06,bou06}.  
We concluded that it is possible to concoct an idealized experiment 
capable of detecting a small fraction of incident gravitons; however, when anything
remotely resembling realistic physics is taken into account, detection becomes impossible.
(Smolin \cite{smo85} has used similar arguments to invoke the necessity of an instrinsic
entropy associated with the gravitational field.)
While one may not conclude that gravitons are in principle undetectable, one can 
conclude that they never will be directly detected.  This leaves Dyson's conjecture in
an ambiguous state.  However, we also found that it is physically impossible, i.e., 
impossible in principle, to detect a given single graviton with resonable probability.  
This latter conclusion has implications for the effect that the emission of gravitation 
radiation might have on quantum coherence as will be discussed in \S \ref{GWs}.  
The following back-of-the-envelope argument for the 
undetectability of a single graviton is typical of these types of analyses.

Consider a resonant-mass gravitational wave (GW) antenna as a protypical GW detector.
The GW cross-section of an undamped harmonic oscillator of mass $m$ and 
length $l$ is approximately \cite{mis73}
\begin{equation}
\sigma \sim \frac{Gml^2 \omega^2 \delta t}{c^3}.
\label{cross}
\end{equation}
where $\omega$ is the resonant frequency (equal to the frequency of the GW), $G$ is the 
gravitional constant, $c$, the speed of light, and $\delta t$, the duration of the harmonic GW.

Now suppose that a gravitational wave pulse is incident on an ensemble of such detectors 
distributed in a sphere of radius $R$.  Further suppose that the incident GW beam is focused
(this is not necessarily even possible for GW's)
such that it has a width of $\sim R$ so as to maximize the flux on the detectors.  Because
$\delta x \sim R$ and $\delta x \delta k \sim 1$, the spread in wavenumber, $k$, of the pulse is
$\delta k \sim 1/R$.  Therefore,  the spread in frequencies is $\delta \omega \sim c/R$.
Then $\delta \omega \delta t \sim 1$ implies a pulse duration of $\delta t \sim R/c$.
Finally, suppose that the total energy of the pulse is that of a single graviton, $\hbar \omega$,
which implies an incident flux, $f$, of 
\begin{equation}
f \sim \frac{\hbar \omega}{R^2 \delta t}.
\label{flux}
\end{equation}

The energy absorbed by a single detector is $f \sigma \delta t$ and the total energy absorbed by all the
detectors is
\begin{equation}
E_{abs} \sim \frac{G M \omega^2 l^2}{c^4 R} \hbar \omega
\label{absorb}
\end{equation}
where M is the total mass of the ensemble of detectors.  This is a classical argument but one can
infer the quantum mechanical interpretation that $E_{abs}/\hbar \omega$ is the probability that a 
quantum of energy is absorbed by any of the ensemble of detectors.  So, in order that this single 
graviton be detected with high probability, $E_{abs} \sim \hbar \omega$.  The 
term $\omega l$ is the order of the speed of sound, $v_s$, in the detector.  Then 
\begin{equation}
(R_S/R)(v_s/c)^2 \sim 1
\label{fraction}
\end{equation}
where $R_S = 2GM/c^2$ is the Schwarzschild radius of the ensemble of detectors.  For ordinary materials, 
$v_s/c \sim 10^{-5}$ and, in any case, $v_s < c$; therefore, the condition for absorbing a significant 
amount of energy from the wave, i.e., detecting a graviton with high probability, is $R < R_S$, 
a condition that cannot be met in principle.
The overall conclusion is that a single graviton cannot be detected
with reasonable probability.  

Of course, the detector envisioned above, a resonant mass, is not
the only conceivable GW detector.  However, it is straightforward to convince oneself that the condition
$R < R_S$ also holds for a Michelson interferometer detector and the conjecture is that
the result is quite general.  One might also consider employing more ensembles of detectors at greater
distances ($r > c\delta t$); however, due to the dispersion of the focused wave the number of additional 
detectors required to intercept the beam actually results in $r/{R_S}^{\prime} < R/R_S$, where 
${R_S}^{\prime}$ is the Schwarzschild radius of the larger system, and so more detectors
only aggravates the situation.

The above argument was offered to address the detectability of a single graviton.  One of the
motivations of the present paper is just the result that it is impossible to directly
detect a graviton and  part of the nonquantum conjecture of this paper is that gravitons, as such,
do not exist.  The above calculation is completely classical and so can also be 
intepreted as indicating that a classical GW pulse with energy less than $\hbar \omega$
cannot be detected.  In \S \ref{GWs} this will be invoked in the context of quantum 
interference.  One might ask whether or not classical gravitational waves 
of arbitrarily small amplitude are possible; however, because such waves are in principle
undetectable, the question is rendered unanswerable and is, therefore, irrelevant.

\section{Gravity and Quantum Interference}
\setcounter{equation}{0} \label{interference}

Quantum effects are nowhere more apparent than in the context of quantum interference and we shall
use the standard two-slit interference experiment as a prototype quantum system.  This
system will be analyzed with back-of-the-envelope/order-of-magnitude methods only, which
is  sufficient for the present purpose.  Two important questions to be addressed are:
1) What conditions must be satisfied in order that the gravitational effects of the system 
can be observed? and 2) If gravity is a nonquantum phenomenon, then how does one treat 
gravity in the context of a coherent quantum system such as a quantum interference 
experiment?  The first of these questions can be simply addressed by considering the
double slit experiment, while the second is open to conjecture. 

Suppose that a nonrelativistic, neutral particle of mass $m$ is in a momentum eigenstate 
with momentum $p$ and is incident on a double slit screen with slit separation $r$.
A neutral test particle of the same mass is located near the screen between the two slits.
The position of the test particle is monitored in order the detect the presence of the
incident particle via its gravitational attraction.  Equal masses for the
two particles turns out to be the optimum choice.  If the position of the incident particle
can be detected with enough precision to determine through which slit it passes, then
presumably the subsequent interference pattern (resulting from an ensemble of 
incident particles detected behind the slits) will be washed out.  

The gravitational acceleration of the test mass, $a_t$, due to the incident particle is
$a_t \sim Gm/r^2$ when the incident particle is near the screen and is insignificant 
when it is far away.  Then, to order-of-magnitude, the net change in the position of
the test particle will be
\begin{equation}
\delta x_t \sim \frac{G m t^2}{r^2}
\label{deltax}
\end{equation}
where $t \sim r/v_i$ is approximately the time the incident particle is near the 
test particle and $v_i$ is the velocity of the incident particle.
The uncertainty in the test particle position is given by the Heisenberg 
uncertainty relation,
\begin{equation}
\Delta x_t > \hbar/\Delta p_t =  \hbar/m\Delta v_t,
\label{heisenberg}
\end{equation}
and a criterion for detection is $\delta x_t > \Delta x_t$.  In addition, $\Delta v_t$ 
must be small enough so that in the time $t$, $\Delta v_t t < \delta x_t$.  Combining
these relations, $\delta x_t > \hbar t/m \delta x_t$.  Substituting $\delta x_t$ from 
Eq. \ref{deltax} yields the following inequality for $t$,
\begin{equation}
t > \frac{\hbar^{\frac{1}{3}} r^{\frac{4}{3}}}{G^{\frac{2}{3}} m}.
\label{time1}
\end{equation}
In order for robust quantum interference to be observed, the deBroglie wavelength of 
the indicident particle, $\lambda \sim \hbar/p_i$ must be on the order of or smaller 
than the slit separation.  Because the velocity of the incident particle is related
to the interaction time by $v_i \sim r/t$, this leads to another inequality,
\begin{equation}
t < \frac{m r^2}{\hbar}.
\label{time2}
\end{equation}
Combining Eqs. \ref{time1} and \ref{time2} yields the following inequalities
\begin{equation}
r > \frac{\hbar^2}{Gm^3},
\label{requation}
\end{equation}
\begin{equation}
t > \frac{\hbar^3}{G^2 m^5},
\label{tequation}
\end{equation}
\begin{equation}
a_t < \frac{G^3 m^7}{\hbar^4},
\label{aequation}
\end{equation}
\begin{equation}
v_i < \frac{G m^2}{\hbar}.
\label{vequation}
\end{equation}
These values should be interpreted as the conditions that must be satisfied for a
gravitational measurement to be made that will sufficiently localize the
incident particle so as to destroy the quantum interference.  If the conditions
are not satisfied, the gravitational interaction is insufficient to detect the
incident particle and quantum interference remains intact.  

There are two other
conditions that should be imposed on this system: 1) the gravitationally
induced motion of the test particle should be less than the slit separation,
$\delta x_t < r$; otherwise, the test particle will not remain within the 
vicinity of the system; and 2) the gravitationally induced motion of the incident
particle should be less than the slit separation, $\delta x_i < r$; otherwise, 
the inequality in Eq. \ref{time2} can be violated and quantum interference compromised.
It can be readily shown that these additional conditions are also consistent with 
Eqs. \ref{requation}, \ref{tequation}, \ref{aequation} and \ref{vequation}, 
and that it is, in fact, these conditions that result in the 
optimal choice of equal masses.  [Note that dimensionless numerical factors have been 
ignored in the above relations, which is consistent with the order-of-magnitude 
philosophy of the analysis.]

None of the above constraints can be used to place specific limits on any of the
parameters, $m$, $r$, $t$, $a_t$, and $v_i$, because each of the relations
involve a different pair of the parameters.  However, one can use Eq. \ref{tequation} 
to limit the mass $m$ by imposing the additional, modest constraint that 
$t$ be less than the age of the universe, $t_u$.  Then
\begin{equation}
m > \frac{\hbar^{\frac{3}{5}}}{G^{\frac{2}{5}} {t_u}^{\frac{1}{5}}}
\label{mequation}
\end{equation}
or $m > 10^7 \, m_p$ where $m_p$ is the mass of the proton.  We conclude that for 
quantum coherent systems with masses less than $\sim 10^7 \, m_p$, there is no
measurable gravitational effect that would compromise their coherence.  The
corresponding limits on the other parameters of interest are
$r > 3 \times 10^3\, cm$, $a_t < 10^{-31}\,cm \,s^{-2}$, and $v_i < 10^{-14}\, cm\,s^{-1}$.
To be sure, $t \sim t_u$ is an extreme limit but because of the large power 
of $m$ in Eq. \ref{tequation}, the limit on $m$ is not overly sensitive to $t$ in an
order-of-magnitude sense.  For example if $t \sim 1s$, the limit on $m$ is
only increased to $3 \times 10^{10} \, m_p$ with $r > 2 \times 10^{-7} \, cm$, 
$a_t < 2 \times 10^{-7 }\, cm\,s^{-2}$, and $v_i < 2 \times 10^{-7}\, cm\,s^{-1}$.  
Even the latter values of $a_t$ and $v_i$ are so extreme that an experiment
capable of detecting them is barely conceivable and so $m > 10^{10} \, m_p$ can
be viewed as a practical lower mass limit.  

>From the small velocities above, one might guess that allowing the incident 
and test particles to be relativistic won't change the results.  This is, indeed, 
the case.  It is straightforward to show that for an ultrarelativistic, 
$\gamma >> 1$, incident particle the net change in the position of the test
particle in the transverse direction is the same as in Eq. \ref{deltax} 
with $t = r/c$, i.e.,
\begin{equation}
\delta x_t \sim \frac{Gm_i}{c^2}
\label{ultra}
\end{equation}
while in the direction parallel to the trajectory of the incident particle the
displacement is 
\begin{equation}
\delta x_t \sim \frac{\gamma G m_i}{c^2}.
\label{ultra2}
\end{equation}
In order to detect either of these motions, clearly $\delta x_t > \ell_{Pl}$ 
where $\ell_{Pl}$ is the Planck length.  Combining this relation with
Eq. \ref{ultra2}, implies $\gamma m_i > m_{Pl}$ where $m_{Pl} \sim 10^{19}\,m_p$ 
is the Planck mass.  So unless $\gamma$ is extraordinarly large, the constraint
on $m_i$ is even more severe than in the nonrelativistic case.  Eq. \ref{ultra}
yields an even stronger constrain, $m_i > m_{Pl}$; however, it's only necessary
to measure the displacement of the test particle in a single direction to detect
the gravitational force of the incident particle.  Using the weaker constraint
$\gamma m_i > m_{Pl}$, the deBroglie wavelength of the incident particle is
\begin{equation}
\lambda_{dB} \sim \frac{\hbar}{\gamma m_i c} < \ell_{Pl}.
\label{debroglie}
\end{equation}
Because the physical size of any particle is much larger than
$\ell_{Pl}$, it is clear that coherent interference in this system cannot be
observed.  Therefore, ultrarelativistic particles, as expected, do not alleviate
the nonrelativistic constraints.

To facilitate comparison with another
condition, note that $t_u \sim 1/\sqrt{G\rho_{crit}}$, in which case 
Eq. \ref{mequation} becomes
 \begin{equation}
m > \frac{\hbar^{\frac{3}{5}} {\rho_{crit}}^{\frac{1}{10}}}{G^{\frac{3}{10}}}.
\label{m2equation}
\end{equation}
where $\rho_{crit}$ is the nominal critical density of the universe.  Because $\rho_{crit}$
appears only to the $\frac{1}{10}$ power in the relation, it can be taken as anything roughly 
of this magnitude, e.g., the density of dark matter, the density of baryons, or even 
the density of cosmic microwave background photons.

There are other conditions that might serve to constrain the mass of such systems.
Because the durations of the experiments are so long and the accelerations so
small, it is likely that environmental noise will place limits
on the measurement.  This is especially true for gravitational environmental noise,
which cannot be shielded.  Suppose there is an unbalanced mass $M$ at a distance
$R$ from the interference system.  Then the differential acceleration between the
the incident and test masses will be $\delta a \sim GMr/R^3 \sim G\rho r$ where
$\rho$ is the mean density of the environmental mass spread over a volume of linear 
dimension $R$.  If we require that $a_t > \delta a$ then Eqs. \ref{requation}
 and  \ref{aequation} imply a mass
limit given precisely by Eq. \ref{m2equation}  with $\rho_{crit}$ replaced by $\rho$.  
(Of course, it might be possible to orient the interference experiment to minimize the 
differential acceleration so long as all the sources of gravity can be
accurately determined.)  As an example, consider the effect of a nearby solar mass 
star at a distance of 10 light years.  The effective, average density of this star
alone is $2 \times 10^{-24}\, g \,cm^{-3}$, which implies a mass limit of
$m > 2 \times 10^7 \, m_p$.  Again, the $\frac{1}{10}$ power of the density results in little 
difference from the previous limit in Eq. \ref{m2equation}.  Note that 
the time scale for an experiment with this mass is $t > 10^8$ years, 
much longer than the star is likely to remain 
within 10 light years.   A more proximal source of gravitational noise
might be a $5000\, kg$ truck passing by $100\, m$ from the experiment.  In this case,
the limit becomes $m > 2 \times 10^9 \, m_p$; although, the duration of the experiment 
relevant to this mass is $> 3$ weeks so that a single passing truck would not compromise
the measurement.  Larger scale motions, such as ground water movement, might do so.
We note in passing that if the acceleration is constrained to be
above the characteristic value in Milgrom's MOND theory of gravity \cite{mil83},
$a_t < 10^{-8}\, cm\,s^{-2}$, then the mass limit becomes $m > 2 \times 10^{10} \, m_p$.

A more insidious type of environmental noise is the background of gravitational
waves that permeate the universe. (The fact that such a background has
yet to be detected is an indication of the level of extreme precision required
for the gendanken experiments considered here.)  Suppose the the gravitational
background at frequency $\omega$ is characterized by a dimensionless metric
perturbation $h$.  The relative acceleration of the incident and test masses is given by
$a_{gw} \sim \omega^2 h r$ and the equivalent mass density of this wave is 
$\rho_{gw} \sim \frac{\omega^2 h^2}{G}$ \cite{mis73}. 
Therefore,
\begin{equation}
a_{gw} \sim \frac{\sqrt{\Omega_{gw}}\, \omega r}{t_u}
\label{agw}
\end{equation}
where $\Omega_{gw} \equiv \rho_{gw}/\rho_{crit}$ and $t_u \sim 1\sqrt{G\rho_{crit}}$.
Combining this expression with the inequalities for $r$ and $a$ in Eqs. \ref{requation} 
and \ref{aequation}, the implied lower limit of $m$ is
\begin{equation}
m > \frac{\hbar^{\frac{3}{5}}\omega^{\frac{1}{10}}\Omega^{\frac{1}{20}}}{G^{\frac{2}{5}}{t_u}^{\frac{1}{10}}}.
\label{m3equation}
\end{equation}
Typical estimates for the equivalent mass densities per octave of the gravitational
wave background range from $10^{-7}\rho_{crit}$ to $10^{-18}\rho_{crit}$ depending on
the frequency; however, because of the $\frac{1}{20}$ power dependence, it barely matters
and for an order-of-magnitude estimate we take $\Omega^{\frac{1}{20}} \sim 1$.
For the lowest possible frequency, $\omega \sim 1/t_u$, the mass limit is 
precisely the same as in Eq. \ref{m2equation}, i.e., $m > 10^7 \, m_p$.  For the much higher
frequency of $\omega \sim 1 \, s^{-1}$, $m > 5 \times 10^8 \, m_p$ about a factor of 60
smaller than the above fundamental limit for $t \sim 1\, s$.

Clearly the shorter the duration of the measurement, the larger the lower mass limit
and it is reasonable to ask if there is a largest value of such a lower limit. 
Setting $v_i = c$ in Eq. \ref{vequation}, yields $m > \sqrt{\hbar c / G} = m_{Pl}$ 
where $m_{Pl}$ is the Planck mass.  In fact all the limits in Eqs. \ref{requation},
\ref{tequation}, and \ref{aequation} become their corresponding Planck values.

The conclusions from these sorts of analyses is that there is a mass limit below which
particles cannot be detected via a gravitational interaction.  Test masses 
below this mass cannot detect a coherent quantum particle and, therefore, cannot frustrate
quantum interference experiments via gravitational measurements/interactions.
There is a fundamental lower limit of $m > 10^7 \, m_p$ for an experiment of the duration
of the age of the universe; however, the lower limit for anything approaching a practical 
experiment is on the order of $10^{10} \, m_p$ and in no case does the lower mass limit 
exceed the Planck mass, $m_{Pl} \sim 10^{19} \, m_p$.  These limits span the mesoscopic 
mass scale.  Even though the values were obtained with rough,
order-of-magntitude estimates, the high power of the mass in the above inequalities
renders the estimates fairly robust.   

The conclusion of the above argument is that the question of whether or not a coherent 
quantum system is the source of a well-defined gravitational field is unanswerable 
for systems with masses $< 10^{10} \, m_p$.  This leaves open the question of 
whether mesoscopic coherent systems are sources of gravity as well 
as questions about the theoretical consistency of
the nonquantum conjecture.  These will be addressed below.
Another, anecdotal piece of evidence for the nonobservability of gravitational interactions 
in certain quantum systems
is provided by the 3d to 1s transition rate in hydrogen due to the emission of
a graviton.  A linear field theoretic calculation of this rate \cite{bou06}
 gives a value of $5.7 \times 10^{-40}s^{-1}$, i.e., the decay time is roughly
$10^{22}$ times the age of the universe, which is again absolutely undetectable.
Before moving on to coherent quantum systems above the mass limit, we next 
investigate the question of whether or not the emission of gravitational radiation 
might frustrate the observation of quantum interference.

\section{Coherence and the Emisson of GWs}
\setcounter{equation}{0} \label{GWs}

There is another way that quantum interference might be frustrated even
in the absence of a particular gravitational measurement and that is if
the diffracting particle emits gravitational radiation.  In that case, 
it might be argued that whether or not there is an observer of the experiment,
relevant information about it, i.e., knowledge of through which slit the particle
passes, is carried to infinity by a gravitational wave.  Because in the
context of this paper GWs are classical phenomena, one might conclude that a
measurement has been performed and, therefore, quantum interference will
be destroyed.  However, what if the emitted radiation is below the fundamental detection 
limit discussed in Section 2?  Then, perhaps, one should consider this as a case 
where no information is transmitted, in which case no measurement has 
been made and quantum interference remains intact.  While this may seem somewhat
inconsistent and/or arbitrary, I remind the reader that, in this paper, only inconsistencies that are
in principle verifiable are considered problematic.  Of course, a quantum theory
of gravity would also predict that no measurement has been made because no
gravitational energy less than $\hbar \omega$ can be emitted from the system.

Returning to the double slit experiment with an incident particle of mass $m$ in a 
momentum eigenstate, it is straightforward to estimate the radiated gravitational
power with the order-of-magnitude relation \cite{mis73}
\begin{equation}
P_{gw} \sim \frac{G}{c^5}(P_{int})^2
\label{power}
\end{equation}
where $P_{int}$ is the internal quadrupole power sloshing around in the emitting system.  
In the present case, $P_{int} \sim mv^2/t$, and the total GW energy radiated is
\begin{equation}
E_{gw} \sim P_{gw} t > \frac{G m^2 v^4}{c^5 t}.
\label{energy}
\end{equation}
In order that the GW be detectible $E_{gw} > \hbar \omega$.  Because
$\omega t \sim 1$ then
\begin{equation}
m > \sqrt{\frac{\hbar c}{G}} \, \frac {c^2}{v^2} > m_{Pl}.
\label{m4equation}
\end{equation}
If one considers the Planck mass as defining the upper mass limit of the
mesoscopic scale, then the conclusion is that the emission of gravitational
waves cannot frustrate coherence in mesoscopic scale systems.  That this
constraint is weaker than those of \S \ref{interference} is, perhaps, not surprising
since the emission and detection of gravitational waves is more
challenging than near field gravitational measurements.  For macroscopic systems,
however, gravitational wave emission would be a source of decoherence.

\section{Gravity and Decoherence}
\setcounter{equation}{0} \label{decoherence}

The conjecture that the gravitational field of a coherent quantum system is not 
well-defined has no
observable consequences for systems with masses $< 10^{10} \, m_p$, as was shown
in \S \ref{interference}.  However, systems above this mass limit can be probed 
gravitationally and so it might be possible to devise an experiment that 
would test the conjecture.  Note, however, that quantum coherence, in the sense we are using
the term here, has not been demonstrated in anything close to such large systems 
\cite{hac03a} and it has 
even been suggested that the random gravitational wave background discussed 
in \S \ref{interference} might lead to an unavoidable decoherence mechanism 
for mesoscopic mass scales \cite{lam06}. Penrose \cite{pen96}, following the ideas 
of Diosi \cite{dio89} and Ghirardi et al. \cite{ghi90}, suggests that there 
is a gravitationally induced spontaneous quantum 
state reduction and that the associated time scale is on the order of 1 
second for mesoscopic scales of the order considered here.  If either of these is the 
case, then an experiment to detect gravitational effects in coherent mesoscopic 
systems becomes even more difficult.

The study of decoherence in quantum systems has been advanced by Zurek \cite{zur81} 
and others in order to understand, entirely within the context of quantum theory, the 
interaction of quantum systems immersed in a surrounding environment.  ``In short, decoherence 
brings about a local suppression of interference between preferred states selected 
by the interaction with the environment'' \cite{sch05}.   Because macroscopic
systems invariably undergo decoherence on very short time scales, they behave
as they would in a classical world, i.e., no quantum interference effects.  One of 
the  main successes of the decoherence program is a precise (quantum mechanical) 
description of how specific quantum mechanical coherent states interact with a (quantum)
measuring apparatus to produce precisely the probability distribution for measurements
that was the core of the Copenhagen interpretation of quantum mechanics.  

One might conclude that decoherence theory has solved the ``measurement problem'' in
quantum mechanics.  This is clearly not the case.  ``What decoherence tells us, is
that certain objects appear classical when they are observed. But what is an observation? 
At some stage, we still have to apply the usual probability rules of quantum theory.''
\cite{jos00}  The probability density matrix may be diagonal in the possible outcomes of
a particular measurement; however, there is no implication that the quantum system realizes
one of those possible outcomes, i.e., there is no implied ``collapse of the wave function''.
While decoherence theory may not satisfactorily explain the transition of a quantum to a 
classical system, one might suspect that when a quantum system becomes decoherent then it 
also becomes a well-defined classical system, whether or not this transition is or ever will 
be understood in terms of a physical theory.  

The transition of a quantum system to a classical system is critical to the conjecture of 
this paper because it is supposed that it is the classical mass distribution that generates 
a well-defined gravitational field.  At least in the macroscopic case, I suspect that nearly 
all physicists would agree with this statement.  For example, I'm sure most would agree that 
it is either one or the other of Schroedinger's
cats that is the source of a gravitational field even before one peers inside the box.
An actual experiment of this effect was performed by Page and Geilker \cite{pag81} and will be
discussed in \S \ref{consistency}.  However, in the case of mesoscopic, decoherent systems I doubt 
that the agreement would be quite so general.  The conjecture of this paper implies that
the gravitational field of any coherent, mesoscopic system is not well-defined but that a 
decoherent, mesoscopic state is the source of a gravitational field, which is, in principle,
measurable.  Such an assertion can lead to inconsistencies, e.g., nonconservation of momentum, 
as will also be discussed in \S \ref{consistency}.  Of course, the same conjecture is made regarding
``microscopic states'' (with $m < 10^{10} \, m_p$); however, there is no measurement that
can confirm or refute it (see \S \ref{interference}).

According to decoherence theory, interactions of an initially coherent
quantum system with its surroundings ``superselect'' a preferred set of orthogonal 
basis states that are stable and correspond to classically observed quantities.  For
macroscopic systems, this superselection process, i.e., decoherence, 
occurs very rapidly.  The interaction Hamiltonians of such systems usually depend on 
position (and other classical quantities) and the resulting preferred basis 
consists of position eigenstates, which typify classical descriptions \cite{sch05}.  
It is precisely this type of preferred basis that is needed to specify the classical
stress-energy tensor, the source of gravity.  In microscopic systems, on the other
hand, energy eigenstates are often the perferred, stable basis and position remains
a property of coherent, non-localized wavefunctions and, therefore, quantum interference 
remains intact \cite{sch05}.

To be sure, there are macroscopic systems that are, in a certain sense, coherent, for
example, superconducting and superfluid systems.  It would seem absurd that 
such macroscopic systems are not sources of a well-defined gravitational field.  While a superfluid Bose 
condensate may be coherent in some respects, it has a classically well specified 
mass density (stress-energy tensor) to which anyone who has observed superfluid helium in a 
glass dewar can attest.  Therefore, superfluids still qualify as classical sources
of gravity.  This behavior is undoubtedly due to the many particle nature of a macroscopic 
superfluid.  Electromagnetic radiation provides another such example.  A classical,
monochromatic electromagnetic wave is a superposition of many coherent photons and
this superposition can certainly exhibit interference.  However, the stress-energy
tensor of  the electromagnetic field is well-defined and undoubtedly acts as a source
of classical gravity (although, this has never been directly observed).  Again, this behavior 
is undoubtedly due to the many particle nature of the system.  So it seems that the
general term ``coherence'', as I have been using it, is not precise enough to describe 
these systems.

Consider the example of a macroscopic crystal that has been cooled
to sub-microKelvin temperatures.  In such a system the electrons are
in a ``coherent'' ground state as is the ion lattice if it is in a
phonon ground state.  On the other hand, the macroscopic crystal can
be extremely well-localized and certainly not in a (coherent) momentum eigenstate 
even if it happens to be moving.  So I'm forced to refine the use of
``decoherent'' and ``coherent'' to refer explicity to whether or not the
the stress-energy, by which we usually mean mass density, of the system is
well-defined, i.e., localized.  That is, a macroscopic body can consist of microscopic
parts, some of which are coherent, and still constitute a decoherent, localized
system.

For mesoscopic scale crystals, it is possible to conceive of a coherent 
superposition of two states, the ground state and first excited phonon state,
and then perform some sort of interference measurement on them that
reveals their quantum coherence.  In fact, Marshall {\it et al.} \cite{mar03} and
Armour {\it et al.} \cite{arm02} have proposed such experiments in which a 
superposition of two different phonon states in a mesoscopic vibrator is probed
with a photon in the former case and a Cooper-pair in the latter.  While such systems
have yet to be realized, even they would not be in the regime to test the 
nonquantum conjecture of this paper.  The mass densities of the systems of the two
superposed states are nearly identical and it not possible to distinguish between
the two states via their (classical) gravitational interaction with a test 
particle without violating the uncertainty relation.  In this example of 
mesoscopic, internally coherent states, the stress-energy tensor is still well
defined enough that classical gravity can prevail.   So it seems that the 
nonquantum conjecture has to be revised to include such cases, i.e., 
``coherent'' superpositions whose mass densities are never-the-less
well localized should also be considered sources of classical gravity.

According to the present nonquantum hypothesis, macroscopic systems are sources of gravity
as described by the classical Einstein's equations while the gravitational fields of 
coherent microscopic systems are not well-defined.  Whether or not 
decoherent microscopic systems are
sources of gravity is irrelevant because gravitational effects in these systems are, in principle,
undetectable.  This leaves the case of mesoscopic systems.  These systems are normally in 
decoherent, localized states and, consequently, are sources of gravity.  However, it {\it might} be possible 
to create a mescoscopic coherent state which, under the current conjecture, would not be
the source of a well-defined gravitational field. 
(Problems with the consistency of these statements are discussed in 
\S \ref{consistency}.)  There is another possible mesoscopic state and that is a partially 
decoherent system.
Partially decoherent microscopic states with masses up to $\sim 10^3 \, m_p$ have been
generated and observed in interference experiments \cite{mya00,hac03b}.  
As decoherence increases, fringe contrast decreases in accordance with
decoherence theory.  How would a nonquantum theory of gravity treat such systems? 
To answer this question requires an interpretation that goes beyond the usual
quantum prescription.  In order to determine whether a given mescoscopic particle is the 
source of a well-defined gravitational field, we must know whether it is 
or is not in a coherent state.  

Consider, again, the double slit diffraction experiment.  Suppose we advance the 
interpretation of a partially decoherent state as a probability $P_c$ that the
mesoscopic particle is in a coherent state and the probability $P_d$ that the particle 
is in a decoherent state, such that $P_c+P_d=1$.  Those particles in the coherent state
will generate the usual diffraction pattern while those in the position eigenstates will
pass through one of the slits or be stopped by the screen.  The net result will be an
interference pattern with decreased fringe contrast.  One then can predict that,  
with probability $P_d$, a particle will exhibit a detectable gravitational field.  
A decoherence analysis, on the other hand, results in a wavefunction that is partially
entangled with the environment, and every particle has the same wavefunction.
As far as quantum theory is concerned, no other interpretation is needed; however, 
neither does the above interpretation imply any difference in the expected 
interference pattern.  Therefore, the interpretation necessary to predict the gravitational
interaction is not in conflict with the usual predictions of quantum mechanics.
It does, however, give a prediction that might well differ from the prediction of a
quantum theory of gravity, i.e., according to this present nonquantum model, some of the
particles will generate a well-defined gravitational field, while others will not.  If 
this prediction is valid, then it might be possible to observe the consequences with
prepared states of mesoscopic particles.  On the other hand, the discussion in 
\S \ref{consistency} indicates that this might not be possible.

After submitting this paper for review, I discovered an intriguing treatment of the 
interaction of classical and quantum systems, the configuration space model of Hall and
Reginatto \cite{hal05}.  In this model the classical and quantum systems are put on
equal footing and both described probabilistically in terms of ensembles on configuration
space.  This description obviates the need for a Copenhagen-type interpretation and may 
be able to provide a natural path to modeling the interaction of quantum systems with
a classical gravitational field.  In addition to conserving probability and energy, the
model allows for back-reaction on the classical system and provides automatic 
decoherence of the quantum system.  Perhaps such a description can lead to a legitimate
model for nonquantum gravity.

\section{Consistency and Experimental Tests of Nonquantum Gravity}
\setcounter{equation}{0} \label{consistency}

The incompatability of the coexistence of quantum and classical fields has been demonstrated
by many people in many different ways \cite{pag81,epp77,ter06,car99,per01,pad02}; however,
``the general question of whether one can consistently couple classical and
quantum systems is a matter of ongoing research....and is not yet resolved.''\cite{car08,hal05}.
Most of these arguments are formal in
nature and proceed by demonstrating the inconsistency of the mathematical formalisms of
quantum field theory with that of a particular description of classical gravity (or some 
other classical field).  In any case, because the nonquantum conjecture of this paper is not 
a formal theory in any sense, it is not surprising that most of these analyses are not 
particularly  relevant to the present case.  However, two such demonstrations of the
necessity of quantum gravity are less formal and offer experimental (gedanken experimental
in one case) evidence to support their claims.

Eppley and Hannah \cite{epp77} considered two gedanken experiments.  In the first a completely classical
gravitational wave of arbitrarily small amplitude is used to ``detect'' the positon of a 
particle (i.e., collapse its wave function) while only imparting to it an arbitrarily 
small momentum impulse.  In this case, both the
particle's momentum and position can be determined to arbitrary accuracy.  Therefore, either
the uncertainty principle or conservation of momentum must be violated.  Considering the
discussion in \S \ref{graviton} on the detectability of GWs, one should be skeptical of such an 
experiment, even a gedanken experiment.  In fact, Mattingly \cite{mat06} has used a similar 
argument to show that such an experiment is, in principle, impossible to conduct. The
other possibility supposes that observing with a GW does not collapse the wave function. 
The relevant gedanken experiment involves the scattering of a classical GW from the
wave function of one of two entangled particles in an Einstein, Rosen, Podolsky \cite{ein35} 
type experiment. If one of the two particles is observed via some nongravitational method, 
then the entangled wavefunction collapses and a distant observer, by observing the
wave function with a GW, would detect this collapse thereby allowing a signal to be
propogated instantaneously over an arbitrarly large distance.  Mattingly \cite{mat06} attacks 
this gendanken experiment on similar grounds.  Albers et al. \cite{alb08} have criticized the
Eppley and Hannah analyses on the grounds that the interaction between the classical gravitational 
wave and the quantum systems was not adequately specified.  They show that a general 
measurement analysis of the coupled quantum/gravitational wave system leads to no inconsistencies.

Page and Geilker \cite{pag81} carried out an actual experiment to test a particular nonquantum
theory of gravity, namely the gravitational field equations given by Eq. \ref{einstein2}.  The
experiment consisted of a Schroedinger's cat type setup in which the positions of the two 
masses in a Cavendish experiment were determined by the result of a quantum decay
process in a radioactive source.  Because the expectation value of the stress-energy operator
was 1/2 the sum of the stress-energy tensors of the masses in the two positions, the
response of the Cavendish experiment would be 1/2 the sum of the two expected classical
responses.  Of course, the observations revealed otherwise as, I have no doubt, any 
physicist would have expected.  

In fact, the ``semi-classical'' theory of gravity, as expressed in 
Eq. \ref{einstein2}, is bothersome for several reasons.  In atomic physics, the semi-classical
treatment of electromagnetic radiation is used to give a plausible account of the interaction 
of electromagnetic radiation with a quantum system (see, e.g., \cite{sch68}).  That it gives the
correct expression for spontaneous emission from quantum transitions in simple atoms is interesting
but not convincing and such analyses can only be justified by a proper quantum field theoretic calculation.
The same has been shown to be true for the spontaneous emission of gravitons from hydrogen \cite{bou06}. 
However, the method breaks down for more complicated systems.   That is, semi-classical treatments
are introduced primarily in order to guess the results of a full quantum mechanical treatment.
Although mathematically well-defined, the semi-classical theory expressed in Eq. \ref{einstein2}
is quite non-physical.  The standard interpretation of the expectation value on the right-hand 
side of this expression is the probability distribution of the outcomes of ``classical'' experiments 
performed on similarly prepared systems.  That one might consider such a probability distribution 
to be a source of gravity seems rather strange.  A more physical way to interpret Eq. \ref{einstein2} 
might be to postulate that the measured gravitational field would be that due to one of the particular 
observed stress-energy tensors occuring with a certain probability.  Of course, this would be fine 
for decoherent systems but would result in the usual contradictions for coherent states.

Analagous to the Page-Geilker experiment\cite{pag81} for Newtonian gravity, Ford \cite{for82} introduced
a hypothetical experiment involving gravitational waves and concluded, not surprisingly, that 
semi-classical gravity results in different predictions than would a quantum theory of gravity.
Because gravitational waves have not been (and probably never will be) generated and detected in the 
laboratory, Ford's analysis is more akin to the gedanken experiments of Eppley and Hannah \cite{epp77}.
In any case, the analysis of Ford and the results of the Page-Geilker experiment are consistent with 
the nonquantum conjecture of this paper, because the assumption is that the source of gravity is 
the classical stress-energy tensor and not the expection value of a quantum mechanical operator. 

Even though the present nonquantum gravity conjecture is, as was pointed out in the introduction, 
by no means a theory, there are still issues of whether it is experimentally consistent
with the rest of physics.  One of the looming issues is conservation of momentum.  Whether
or not a particle in a coherent quantum state is the source of well-defined gravitational field, 
it is certainly true that the 
gravitational field of a decoherent, macroscopic body interacts with such a particle.
The equivalence principle demands it.  Pound and Rebka's {\cite{pou60} detection of the gravitational
blue shift of gamma rays provides experimental verification.  In that case, a coherent
quantum particle (photon) gains momentum (and energy) from the gravitational interaction with the earth.
However, if our nonquantum conjecture is valid, there is no well-defined gravitational attraction of 
the earth by the quantum particle and thus the momentum impluse received by the earth
can not be determined.  The implication is 
that the total momentum of the system might not be conserved.  Of course, in the Pound-Rebka experiment
this has no observational consequence because measuring such a small change in the earth's momentum
is impossible, as would be the case for any macroscopic system.  It is necessary to 
look for the effect in microscopic or possibly mesoscopic experiments.  

One can show that precisely the same
detection criteria (Eqs. \ref{requation} - \ref{vequation}) 
apply to this case where again the optimum experiment
would employ equal mass particles.  Therefore, it is impossible in principle to perform such
an experiment with particles less massive than $10^7 \, m_p$.  In anything remotely approaching
a realistic experiment this mass limit is surely much greater, perhaps $10^{10} \, m_p$ or even
larger.  On the other hand, if the particle is too massive it will be virtually impossible to
prepare it in a coherent state.  Therefore, a mesoscopic scale experiment (e.g., from 
$10^{10} \, m_p$ to $10^{15} \, m_p$) is the most likely arena.  

Even if an experiment to test
this example of nonconservation of momentum is wholly impractical, i.e., it is relegated to
the realm of a gedanken experiment, it should still be taken as a serious problem for the
present nonquantum conjecture.  On the other hand, if such a measurement were performed on
a coherent particle, the process would be expected to promote decoherence, i.e., wavefunction
collapse, of the coherent particle, in which case the particle would be a source of a well-defined
gravitational field and momentum would then be conserved.  Our
current nonquantum conjecture has nothing to say about how this process would occur.
A standard decoherence analysis would necessarily have
to consider one of the particles as a classical source of a classical field, an
anathema for an inherently quantum mechanical analysis.  On the other hand, the configuration
space model of Hall and Reginatto \cite{hal05} seems to be designed for just this sort of system.
An obvious next step would be to introduce a specific configuration space interaction 
Hamiltonian in order to predict how this transition might occur and determine whether or not
there are observable consequences that might test the model. 
In any case, if gravity is a nonquantum 
phenomenon, it seems likely that it would make itself known on mesoscopic scales.
Salzman and Carlip \cite{sal06} suggested that that experiments on somewhat smaller mesoscopic mass scales 
might be able to test the version of nonquantum gravity expressed in Eq.\ref{einstein2}.  Also, evidence 
for a variety of models of gravitationally induced wavefunction collapse would most likely appear
on mesoscopic scales \cite{dio89,pen96,lam06,van08}.

\section{Discussion and Futher Speculation}
\setcounter{equation}{0} \label{discussion}

It was pointed out in the Introduction that the nonquantum gravity conjecture introduced
in this paper is tantamount to heresy.  Actually, the real heresy is, perhaps, the 
characterization of nonquantum, classical states of matter as fundamentally {\it real}
and legitimate concepts in a fundamental theory of physics.
The tremendous success of quantum theory in the last
80 years has been such that it seems inconceivable that any fundamental theory will not be
quantum in nature.  To be sure, the 1 part in $10^{12}$ agreement of the quantum 
electrodynamic prediction with observations of the gyromagnetic ratio of the electron
is a spectacular confirmation of the theory.  However, this is a single example in 
a specific microscopic system.  Classical electrodynamics also has spectacularly 
confirmed predictions.  For example, Maxwell's equations predict the $1/r^2$ dependence
of the electric field of a point charge and this has been confirmed to an accuracy of
1 part in $10^{16}$.  Still, I suspect that it is the consensus among physicists that classical 
physics is simply an approximate limit to a fundamental quantum theory.  The irony of this view 
is that aspects of classical physics are absolutely necessary in order to give meaning 
to quantum theory.  After all, the predictions that quantum theory makes are of the
statistical outcomes of measurements and these measurements are ultimately described 
in terms of classical physics.  The once orthodox and, perhaps, currently disfavored 
Copenhagen interpretation of quantum mechanics is, in brief, that quantum theory provides
a complete account of microscopic phenomena by making probabilistic predictions of
the outcomes of experiments that are described operationally (i.e., classically).
(Of course, any brief statement of the Copenhagen interpretation is necessarily 
incomplete. \cite{sta72})  It is the accumulated empirical knowledge of how to
(classically) prepare a quantum system and then how to (classically) conduct a
measurement that is essential to give meaning to the theoretical predictions of 
quantum theory.  

I find it rather curious that there is a perceived great need to unify all
the fundamental forces into an all encompassing quantum theory of nature while
the need to unify classical (experimental) physics with quantum (theoretical) 
physics seems much less important
even though the former is absolutely essential in the interpretation of the latter.
It is for this reason that I am willing to consider the classical stress-energy tensor
as the fundamental source of gravity.
To be sure, a primary goal of the decoherence program is to illuminate the 
interactions of quantum systems with measuring apparatus.  This program, by and in
large, has successfully demonstrated how the laws of quantum theory lead to decoherence 
and consequently demonstrated that the resulting probability distributions of outcomes 
of experiments conform to those of classical physics.  However, as to just why it is
that an essentially quantum mechanical world appears classical to us, decoherence theory
is silent.  In the case of Schroedinger's cat, quantum decoherence is able to deomonstrate,
in principle, why it is that the two states of the cat cannot be made to
exhibit quantum interference.  However, as to which of the two states actually occurs
or even that only one of the two states does occur, we have to resort to the usual
probabilitistic interpretation of quantum theory, which links quantum wavefunctions with
classical observations.  I suspect no physicist would doubt that it is the {\it real}
cat, alive or dead, that is the source of a gravitational field, whether or not
an outside observer determines the state of the cat.

I suppose that a legitimate criticism of the conjecture put forward in this paper is its 
lack of predictive power.  Except possibly in the case of the coherent to decoherent
transitions in mesoscopic systems, and even in this case the conjecture makes no 
specific prediction, the nonquantum conjecture makes no additional predictions that
can not already be made by quantum theory and general relativity.  However, even the
simple conjecture in this paper does, in a sense, make the prediction that none of
the specific predictions made by any quantum theory of gravity will be confirmed 
experimentally.  A specific criticism might be that the nonquantum conjecture has
nothing to say about the Planck scale, at which surely some interesting new physics
must appear.  This may be so; however, the Planck density exceeds nuclear densities
by  a factor of nearly $10^{80}$ and I personally have no confidence that any current theory of
physics is valid at this scale.  At a slightly larger scale there are questions associated
with cosmological inflation.
Here again I must admit that the current nonquantum conjecture has nothing to offer; however, even at 
this scale, physics in general is not yet well understood.  

A related issue is that of the singularities predicted by general relativity.  If gravity
is fundamentally classical, will these persist?  If so, then clearly it would be a crisis
for fundamental physics.  The current understanding is that such singularities 
become resolved at the Planck scale by quantum gravity.  As I indicated above, I doubt that
any of our current physical theories, including general relativity, are valid at this scale.
There is currently no experimental evidence regarding such small scales with the possible exception
of inflation in the early universe and indirect observations of that epoch are extremely
limited.  So, at least for now, I'm willing to ignore the singularities implied by general
relativity until more is known (observationally) about the extreme condidtions in their 
vacinities. 

There are also quantum issues having to 
do with both particle creation by and the entropy of black holes \cite{bek73,haw75}.  While these may, 
indeed, be important problems for the consistency of theoretical physics, there is, as far as I
know, no experimental observations relevant to them nor is there even strong observational
evidence that general relativity provides an accurate description at the black hole event horizon.
Nevertheless, the ``information loss paradox'' has garnered a great deal of attention and is
considered by some to be key to our understanding of fundamental physics \cite{pre92,haw05,hay07}.
The paradox arises because the ``no hair'' theorem of classical gravity implies that information
is lost in black holes.  If this is so, then Hawking evaporation of black holes implies that 
pure quantum states evolve into mixed states with the implication that quantum gravity is not
unitary.  However, if gravity is a classical field then, perhaps, nonunitarity is not so strange.
A similar problem occurs in the interaction of a quantum system with a classical measuring apparatus.
The outcome of a particular measurement of such a system is also not consistent with unitary evolution.

The accelerated expansion of the universe is a model with some observational support \cite{bah99} 
and certainly is in desperate need of explanation.  However, it is still possible that
these observations are explained within the context of general relativity 
\cite{kol06} or some classical variant of it.  Finally, there is the ``cosmological constant''
problem, i.e., the problem that quantum theory seems to quite generally imply the
existence of a cosmological constant that is more than $10^{120}$ times larger than
that observed.  One might hope that some future quantum theory of gravity will explain
this.  While the current nonquantum conjecture does not address this problem directly,
if the vacuum is, indeed, a non-localized, coherent state, then the nonquantum conjecture would 
imply the resulting cosmological constant is not a well-defined source of gravity and indicates
that a nonquantum theory of gravity might help solve this problem.
Although, I wouldn't call it a prediction, the
present nonquantum conjecture does suggest that relevant experimental evidence might
appear in coherent and partially decoherent systems of mesoscopic scale ($> 10^{10}\, m_p$)
as discussed in \S \ref{consistency}.

How one might go about incorporating the nonquantum conjecture into a more complete model or 
theory is not clear to me.   The transition from coherence to decoherence, especially the
configuration space model of Hall and Reginatto \cite{hal05}, might offer some clues.
Of course experimental evidence of the gravitational effects of this transition would be 
invaluable; however, both the preparation of such systems and the measurement of their gravitational 
interactions may be virtually impossible by practical standards.

There are two great field theories of classical physics, gravitation and electromagnetism.
I have argued that the source of gravity, i.e., the right-hand side of Einstein's equations,
is the classical stress energy tensor.  Likewise, the source terms of electricity and
magnetism, the right-hand side of Maxwell's equations, are classical charge and current densities.
In the case of gravity, I claim that coherent (non-localized) mass-energy distributions do not
generate well-defined gravitational fields. 
Why not make the same claim for quantum sources of electromagnetism?
The fact is that when microscopic phenomena are probed, one finds evidence of the quantum 
nature of electromagnetism.  The resulting theory of quantum electrodynamics (QED) appears to
describe all electromagnetic phenomena, microscopic and macroscopic; although, the application 
of QED to complicated macroscopic phenomena is problematic at best.  One might argue by analogy that
the general relativity is simply the classical limit of a quantum theory of gravity.  
Arguments by analogy, while often compelling, are also often wrong.
It is the contention of this paper that when one probes deeply into microscopic gravitational 
phenomena, there is simply nothing there.

If the other fundamental forces of nature are described by a unified quantum theory,
why should gravity lie outside this framework?  Certainly, gravity is distinct in 
several respects.  Gravity is extraordinarily weaker than the other 
fundamental interactions and couples universally to all forms of energy.
It also has an inherently global aspect to it, i.e., locally the effects of gravity vanish
in a local inertial (free-falling) frame.  It is the other, Lorentz invariant forces of nature that 
fix the local Minkowskian structure of space-time and it is the theory of gravity, 
general relativity, that tells us how to stitch these local Minkowskian patches together
into a global structure.   In this sense, gravity is associated with the global structure 
of space-time, the stage upon which the other fundamental forces play.  Clearly, this
picture must break down at the Planck scale.  The energy of a Planck frequency photon will, 
upon detection, be localized within it's Schwarzschild radius whether or not gravity is a 
quantum phenomenon.  

Still, it may well be 
that an elegant and useful quantum theory of gravity will be discovered in the future.  
How far in the future, I don't know.  I'm tempted to borrow a statement from Freeman Dyson
who once told me, in a response to my query about string theory, that he thinks string theory
is probably correct, it is simply premature.  How premature, I asked.  About a hundred 
years, he responded.  Perhaps one of the reasons quantum gravity might be premature is 
the current total lack of any observational evidence of the quantum nature of gravity.
As an experimentalist, I think of theories as models that are created to make sense
out of our observations of nature.  Successful theories make additional predictions that are 
then confirmed.  It seems to me that the current quest for a quantum theory of 
gravity is the search for a consistent mathematical model in the absence of
experimental evidence with the hope that the model makes predictions that will someday be
confirmed.  

Striving to comprehend our world is human nature and it is understandable that scientists, 
physicists in particular, relish elegant and simple universal laws that describe the cosmos.
What could be more elegant than a unified  theory of all the fundamental forces 
from which all else follows, especially if that theory were expressed in terms of a 
single physical quantity, for example, the tension in string theory?   However, it may simply be 
that the world is not so tidy and will ultimately elude our attempts to ever more simplify 
our description of it.  Whatever our fundamental theory of the universe might be, its 
stature will be due, in large part, to the experimental observations that support it.  
Let me end with a caution offered by L\'eon Rosenfeld \cite{ros63}.  ``There is no denying 
that, considering the universality of the quantum of action, it is very tempting to regard 
any classical theory as a limiting case to some quantal theory.  In the absence of 
empirical evidence, however, this temptation should be resisted.''
\\[2mm]
{\bf Acknowledgements}

I would like to thank Freeman Dyson for inspiring this work and also Jim Peebles and Tony
Rothman for helpful discussions (and for listening to my crazy ideas).  Marcel Reginatto,
Steve Carlip, Jeff Kuhn, and Bruce Partridge made useful comments on an earlier draft of this
document.

{\small

\end{document}